\documentclass[aps,pra,nofootinbib,preprint,amsmath,amssymb,floatfix]{revtex4-2}
\usepackage{color}
\usepackage{graphicx}
\DeclareMathOperator{\Li}{Li}

\begin{document}

\newcommand{\tc}{\textcolor}
\newcommand{\g}{blue}
\title{Axionic and nonaxionic electrodynamics in plane and circular geometry}
{\author{ Iver Brevik  }      
\affiliation{Department of Energy and Process Engineering, Norwegian University of Science and Technology, N-7491 Trondheim, Norway}
\author{Amedeo M. Favitta}
\affiliation{ Department of Physics, University of Palermo, I-90123 Palermo, Italy  }
\author{Masud Chaichian}
\affiliation{Department of Physics, University of Helsinki, and Helsinki Institute of Physics,  P. O. Box 64, FI-00014 Helsinki, Finland}

\date{\today}          

\begin{abstract}

Various aspects of axion electrodynamics in the presence of a homogeneous and isotropic dielectric medium are discussed.  1. We consider first the  "antenna-like" property of a planar dielectric surface in axion electrodynamics, elaborating on the treatment given earlier on this topic by  Millar {\it et al.} (2017). We calculate the electromagnetic energy transmission coefficient for a dielectric plate, and compare with the conventional expression in ordinary electrodynamics. 2. We consider the situation where the medium exterior to the plate, assumed elastic, is "bent back" and glued together, so that we obtain a circular dielectric string in which the waves can propagate clockwise or counterclockwise. As will be shown, a stationary wave pattern is permitted by the formalism, and we show how the  amplitudes for the two counterpropagating waves can be found.  3. As a special case, by omitting axions for a moment,  we analyze the Casimir effect for the string, showing its similarity as well as its difference with the Casimir effect  of a scalar field for a piecewise uniform string  (Brevik and Nielsen 1990). 4.  Finally, including axions again we analyze the enhancement of the surface-generated electromagnetic radiation near the center of a cylindrical haloscope, where the interior region is a vacuum and the exterior region a high refractive index medium.  This enhancement is caused by the curvature of the boundary, and is mathematically a consequence of the behavior of the  Hankel function of the second kind for small arguments.  A simple estimate shows that enhancement may be  quite significant, and  can therefore be  of experimental interest.
The  presence of an absorber in the center and the possibility of adopting it to search for axions with mass in the THz region, and possibly the GHz region  too, is also discussed.
This proposal is suggested as an alternative to the reflector arrangement in a similar arrangement recently discussed by Liu {\it et al.} (2022).

\end{abstract}
\maketitle

\bigskip
\section{Introduction}
\label{secintro}

The "antenna-like" behavior of a single dielectric plane surface in axion electrodynamics in the presence of a strong external magnetic field is a rather remarkable phenomenon. In particular, it is quite perplexing  that the Poynting vector normal to  a dielectric surface is no longer continuous across the surface; a property otherwise considered as a cornerstone in electromagnetic theory. There occurs an extra electromagnetic radiation of energy from the surface. Of course, this energy cannot stem from nowhere. It is rather a consequence of the interchange of field energies happening between two reservoirs, namely the usual  electromagnetic reservoir, and the other one due to the axion field.  A detailed exposition of these properties of axion electrodynamics is given by Millar {\it et al.} \cite{millar17}. The effect is also planned use of in experimental broadband tests \cite{liu22}.

Our  purpose with the present paper is first to elaborate further on some of the consequences of this dielectric-surface  effect.  After giving a brief survey of  axion electrodynamic formalism in the next section, we consider in Sec. III  the most simple dielectric system where two surfaces are involved, namely the planar dielectric disk.  We point out the close similarity between the transmission coefficients in the axion case and in the non-axion case, a similarity that is not quite trivial.  Thereafter in Sec. IV, still dealing with two-surface systems, we turn the exterior medium 2 "back" and join it with  medium 1 so that we obtain a circular loop or string,  and investigate if the axion  formalism admits a stationary state implying clockwise and counterclockwise propagating modes. Actually the formalism does so.  Such a system is most likely only of fundamental interest, but the positive outcome of the analysis  demonstrates  the flexibility of the axionic formalism. As a byproduct, if we for a moment ignore the axion field we  can calculate the electromagnetic Casimir energy of the system. This is considered in Sec.~V.  We show  how the systems bears similarities to the  piecewise uniform relativistic string which has  been considered repeatedly earlier \cite{brevik90,li91,elizalde93,brevik96,brevik99,brevik02,bordag12}. Finally, in Sec. VI we return to the antenna-like property of dielectric surfaces, and investigate the enhancement of the electric field near the center of a cylindrical haloscope caused by the curvature of the emitting boundary. Here the central region is a vacuum, while the outer region is a metal. The enhancement can actually be quite significant. This opens for the possible measurement of the emitted field from the cylindrical boundary, and may thus be an alternative to the broadband solenoidal haloscope recently proposed by Liu {\it et al.}~\cite{liu22}.

  Some of the pioneering papers on axion electrodynamics are listed in Refs.~\cite{peccei77,peccei77a,weinberg78,sikivie83,preskill83,abbott83,dine83,sikivie08,sikivie14}. More recent works can be found in Refs.~\cite{lawson19,kim19,qingdong19,sikivie03,mcdonald20,chaichian20,zyla20,arza20,carenza20,leroy20,brevik20,brevik21a,oullet19,arza19,
qiu17,dror21,brevik22a,fukushima19,tobar19,bae22,adshead20,patkos22,tobar22,derocco18,brevik22}.

\section{Basic formalism in a dielectric environment }

We consider a pseudoscalar axion $a= a({\bf r}, t)$ present in the entire universe, making a two-photon interaction with the electromagnetic field. We assume  a dielectric environment where the permittivity is $\varepsilon$ and the permeability is $\mu$, where these material parameters are constants. The constitutive relations are  $\bf{D}=\varepsilon {\bf E},  {\bf B}=\mu \bf H$. There are two field tensors,  $F_{\alpha\beta}$ and  $H_{\alpha\beta}$, where $\alpha$ and $\beta$ run from 0 to 3. We assume the standard Minkowski space with the convention $g_{00}=-1$. The dual is defined as $\tilde{F}^{\alpha\beta}=\frac{1}{2}\varepsilon^{\alpha\beta\gamma\delta}F_{\gamma \delta}$, with $\varepsilon^{0123}= 1$.

 For convenience we give the expression for the field tensors explicitly,
\begin{equation}
F_{\alpha\beta}= \left( \begin{array}{rrrr}
0    &  -E_x   & -E_y  &  -E_z \\
E_x  &    ~0     & B_z   & -B_y  \\
E_y  &  -B_z   &  ~0    &  B_x \\
E_z  &   B_y   &  -B_x &   ~0
\end{array}
\right), \label{2}
\end{equation}
\begin{equation}
H^{\alpha \beta} = \left( \begin{array}{rrrr}
0    &  D_x    &  D_y    &  D_z  \\
-D_x &  0      &  H_z   &  -H_y  \\
-D_y &  -H_z    &   0     &  H_x  \\
-D_z &  H_y   &   -H_x   &    0
\end{array}
\right). \label{2a}
\end{equation}

The Lagrangian is \begin{equation}
{\cal{L}}= -\frac{1}{4}F_{\alpha\beta}{H}^{\alpha\beta} +{\bf A \cdot J}-\rho \Phi -\frac{1}{2} \partial_\mu a\partial^\mu a-\frac{1}{2}m_a^2a^2       - \frac{1}{4}g_\gamma \frac{\alpha}{\pi}\frac{1}{f_a}a(x) F_{\alpha\beta}\tilde{F}^{\alpha\beta}. \label{1}
\end{equation}
Here, $\rho $ and $\bf J$ are the usual electromagnetic charge and current densities (since the axions are electrically neutral they can not contribute);
 $g_\gamma$ is a model-dependent constant for which we adopt the value   $ 0.36$  \cite{sikivie03};  $\alpha$ is the fine structure constant, and $f_a$ is the axion decay constant whose value  is
insufficiently known.   Often it is assumed  that $ f_a   \sim 10^{12}~$GeV.

Defining the  combined axion-two-photon coupling constant  as
\begin{equation}
g_{a\gamma\gamma}= g_\gamma \frac{\alpha}{\pi}\frac{1}{f_a},
\end{equation}
we see that  the last term in the Lagrangian (\ref{1}) can be written as ${\cal{L}}_{a\gamma\gamma} =  g_{a\gamma\gamma} a(x)\,{\bf E\cdot B}.$

It is convenient to define the quantity $\theta(x)$,
\begin{equation}
\theta(x)= g_{a\gamma\gamma}a(x).
\end{equation}
Based on the  expression (\ref{1}), the extended Maxwell equations can then be written as
\begin{equation}
{\bf \nabla \cdot D}= \rho-{\bf B\cdot \nabla}\theta, \label{5}
\end{equation}
\begin{equation}
{\bf \nabla \times H}= {\bf J}+\dot{\bf D}+\dot{\theta}{\bf B}+{\bf \nabla}\theta\times {\bf E}, \label{7}
\end{equation}
\begin{equation}
{\bf \nabla \cdot B}=0, \label{8}
\end{equation}
\begin{equation}
{\bf \nabla \times E} = -\dot{\bf B}. \label{9}
\end{equation}
These equations are  general, i.e., there are    no restrictions so far  on the spacetime variation of $a(x)$. The equations are moreover relativistic covariant, with respect to shift of the inertial system.

The governing equations for the fields can correspondingly be written as
\begin{equation}
\nabla^2 {\bf E}-\varepsilon\mu \ddot{\bf E}=    {\bf \nabla (\nabla \cdot E)}
 +\mu \dot{\bf J}+ \mu \frac{\partial}{\partial t}\left[\dot{\theta }{\bf B}+ {\bf \nabla}\theta{\bf \times E}\right], \label{10}
\end{equation}
\begin{equation}
\nabla^2 {\bf H}-\varepsilon\mu \ddot{\bf H}= -{\bf \nabla \times J}-{\bf \nabla \times }[\dot{\theta}{\bf B}+{\bf \nabla}\theta{\bf \times E}]. \label{11}
\end{equation}
The dynamical field entities are here the electromagnetic fields; we limit ourselves to a perturbative approach in which the axions are the perturbation.  We do not consider the field equations for the axions explicitly.

Assume now that a strong static magnetic field ${\bf B}_e
 = B_e\hat{ \bf z}$ acts in a region where  $\rho$ and $\bf J$ are zero and the axion field is spatially uniform but varies harmonically in time,
\begin{equation}
a(t)= a_0\cos \omega_at.
\end{equation}
This is the situation usually found in the inner region of a haloscope. Then it is convenient to separate out the part of $\bf E$ that is caused by the uniformly fluctuating axions. Calling this contribution ${\bf E}_a(t)$, we see from Eq.~(\ref{10}) that it is connected by the $\ddot{\theta}$ term. From the governing equation for ${\bf E}_a(t)$,
\begin{equation}
\nabla^2 {\bf E}_a-\varepsilon\mu \ddot{\bf E}_a=    \mu \ddot{\theta}{\bf B}_e \label{ 14a}
\end{equation}
we then obtain, after omitting the $\nabla^2$ term, the solution
\begin{equation}
{\bf E}_a(t)= -\frac{1}{\varepsilon}E_0\cos \omega_at \,\hat{\bf z},
\end{equation}{
where
\begin{equation}
E_0 =\theta_0 B_e.
\end{equation}
After the separation of  the component ${\bf E}_a$, the field equation (\ref{10})  takes the reduced form
 \begin{equation}
\nabla^2 {\bf E}-\varepsilon\mu \ddot{\bf E}=  {\bf \nabla (\nabla \cdot E)} +\mu {\dot{ \bf J}}+  \mu [ \dot{\theta}\dot{\bf B} + {\bf \nabla}\theta{\bf \times \dot{E}}]. \label{12}
\end{equation}
 Likewise for the magnetic field
\begin{equation}
\nabla^2 {\bf H}-\varepsilon\mu \ddot{\bf H} = -{\bf \nabla \times J}-\left[   \dot{\theta}{\bf \nabla \times B}
 + ({\bf \nabla}\theta){\bf \nabla \cdot E}
 -({\bf \nabla}\theta \cdot {\bf \nabla})  {\bf E}      \right]. \label{13}
\end{equation}

\section{Antenna-like behavior: two plane parallel surfaces}

We begin by considering one single planar dielectric surface, placed at $x=0$, separating the left region 1 (refractive index $n_1$) from the right region 2 (refractive index  $n_2$). We assume the simple case where the media are nonmagnetic and $n_1$, $n_2$  constants and real. A strong static magnetic field ${\bf B}_e=B_e\hat{\bf z}$ is imposed in the $z$ direction.  An incoming wave polarized in the $z$ direction comes in from the left, propagates in the $x$ direction,  and becomes partly reflected by the surface. The basic property following from the extended Maxwell equations in the present context is that the components of $\bf E$ and $\bf H$ parallel to the surface have to be continuous  (as in ordinary electrodynamics) at $x=0$.

This is the situation analyzed in detail by Millar {\it et al.} \cite{millar17}. Because of the axions there will be produced two outgoing electromagnetic waves, one going to the left and one going to the right. In this sense we can consider the dielectric surface to have "antenna-like" properties. It does not mean that electromagnetic energy comes out from nothing, however, but that there occurs an energy interchange  between the ordinary electromagnetic and the axion reservoirs.

It is now convenient, following Ref.~\cite{millar17}, to distinguish the produced traveling fields by an extra index   $\gamma$. Continuity of
 ${\bf E}_\parallel$ gives straightaway
\begin{equation}
E_1^\gamma +E_1^a= E_2^\gamma +E_2^a.
\end{equation}
Next, we may take into account the relationship
\begin{equation}
H^\gamma = \pm \frac{1}{n}E^\gamma, \label{gamma}
\end{equation}
which holds for a propagating wave when the medium is nonmagnetic. This implies that the boundary condition for ${\bf H}_\parallel$  can be written as
\begin{equation}
-n_1E_1^\gamma = n_2E_2^\gamma,
\end{equation}
expressing that the two wave vectors ${\bf k}_1$ and  ${\bf k}_2$  are antiparallel. As $E^a= -(1/\varepsilon)E_0$, we can then solve for the produced fields to get
\begin{equation}
E_1^\gamma= -\frac{E_0}{n_1}\left( \frac{1}{n_2}-\frac{1}{n_1}\right), \quad E_2^\gamma= \frac{E_0}{n_2}\left( \frac{1}{n_2}-\frac{1}{n_1}\right), \label{check}
\end{equation}
(recall that $E_0=\theta_0B_e$).

Remark: The argument above did not imply there to be any infalling initial photons from the left. This is the most fundamental constellation. The argument however implies a nontrivial point, namely the use  of Eq.~(\ref{gamma}) in an initial vacuum. This equation is not a general one in electrodynamics, but is implicitly resting on the assumption that there are propagating waves. This point thus brings some concern about the validity of the argument in this special case where there initially a vacuum. Now, Millar {\it et al.} discuss also the more general case where there are infalling photons from the left, and in that case the use of Eq.~(\ref{gamma}) is of course nonproblematic.

Now move on to consider the situation with two dielectric surfaces; we focus first on simple dielectric slab of thickness $d$,surrounded by  vacuum. The refractive index is $n=\sqrt{\varepsilon}$, similarly as above. It is of interest to consider the energy transmission coefficient
\begin{equation}
T= \big|\frac{E_T}{E_I}\big|^2
\end{equation}
in the presence of axions,
 $E_T$ and $E_I$ referring  to the transmitted and incident wave amplitudes. Based upon Eq.~(5.4a) in Ref.~\cite{millar17} we obtain
 \begin{equation}
 T= \frac{4n^2}{4n^2+ (n^2-1)^2\sin^2 kd},
 \end{equation}
 where $k= 2\pi/\lambda= n\omega  $. Remarkably enough, this expression does not contain $E_0$. Some special cases are noteworthy: if $d=\lambda/4$ the transmission is at minimum, $T= T_{\rm min}= 4n^2/(n^2+1)^2$, whereas if $d=\lambda/2$, the transmission is at maximum, $T =T_{\rm max}=1$.

 It is of interest to compare this transmission coefficient from that occurring in ordinary electrodynamics. Giving the latter quantity an extra subscript elmag, we obtain (page 514 in Stratton \cite{stratton41})
 \begin{equation}
 T_{\rm elmag}= \frac{4n^2}{(n^2+1)^2-(n^2-1)^2\sin^2 kd}.
 \end{equation}
It is seen that $T$ and $T_{\rm elmag}$ are different, what could be expected since their derivations are different. Now, if $d=\lambda/4$, $T_{\rm elmag}= 1$, while if $d=\lambda/2$,
$T_{\rm elmag}= 4n^2/(n^2+1)^2$, its minimum value.
{
It ought to be mentioned that the reflection and transmission coefficients refer to the axion-generated fields only. If there were external probe source fields, $E_{\rm probe}$,  present, then its effect would be suppressed by a factor $(E_0/E_{\rm probe})^2$.}

\section{The closed string geometry}

In this section we will continue to consider dielectric systems containing two interfaces separating media of refractive indices $n_1$ and $n_2$, but now in a form that has not been  considered before as far as we know. Let $n_1$ refer to the left medium, $n_2$ refer to the second medium, and assume that the media are elastic so that medium 2 can be turned back and glued to the left side of medium 1. Therewith we obtain a ring-formed system. One may ask: does such a system allow stationary oscillations to occur when  axions are present? It is of physical interest to examine this point, not because of the applicability of the formalism in practice, but rather as a test of the flexibility of this  kind of generalized electrodynamics.
\begin{figure}[h]\label{fig:1}
	\includegraphics[scale=0.3]{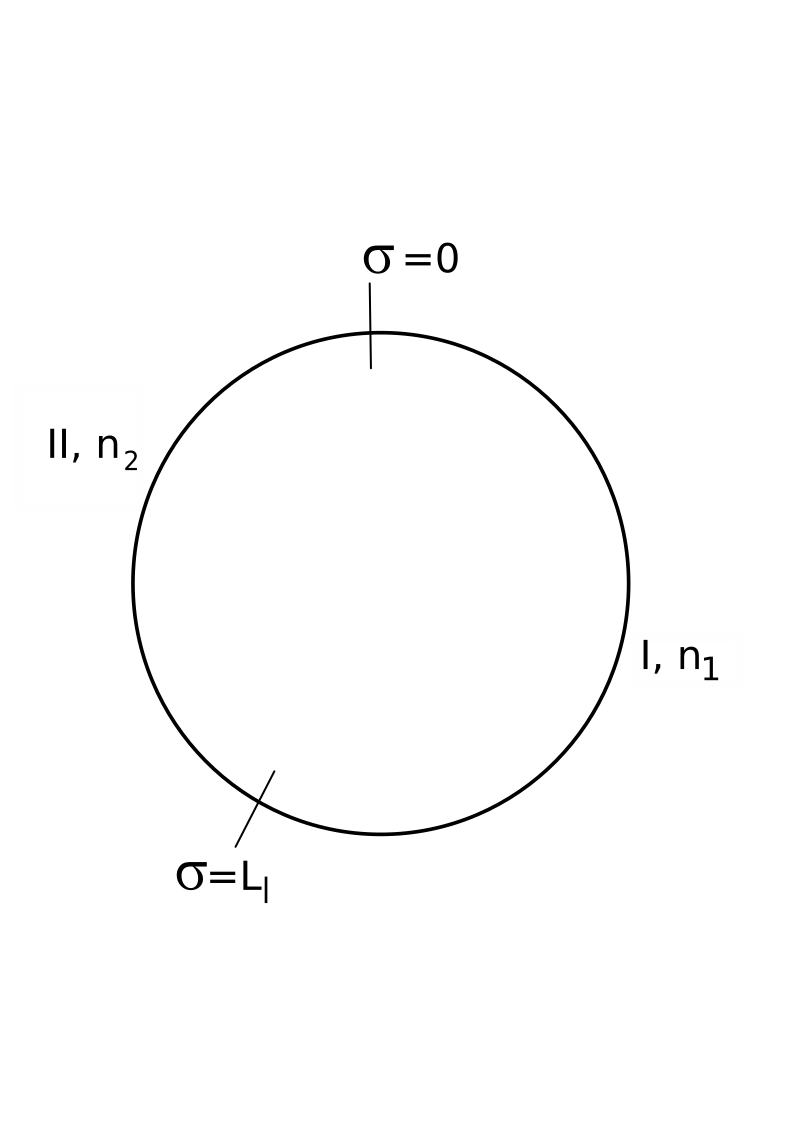}
	\caption{Geometry and notation of the closed string.}
\end{figure}
Figure 1 shows the configuration. We let $\sigma$ denote the length coordinate along the string, such that the two dielectric junctions are at $\sigma=0$ and $\sigma = L_{I}$. The total length of the string is $L=L_I+L_{II}$, so that the junctions $\sigma=0$ and $\sigma=L$ are overlapping. We will be interested in the fields in the interior regions of the string. The string is lying in the $xy$ plane, and a strong uniform magnetic field ${\bf B}_e$ is applied in the $z$ direction.

We will seek for stationary oscillations of the electromagnetic oscillations in the string. If $E_I(\sigma,t)$ and $E_{II}(\sigma,t)$ are the electric fields in the two regions, we have in complex representation \cite{brevik90}
\begin{equation}\label{25}
E_I(\sigma,t)=e^{-i\omega t}[ \xi_Ie^{in_1\omega \sigma}+\eta_Ie^{-in_1\omega \sigma }],
\end{equation}
\begin{equation}\label{26}
E_{II}(\sigma,t)=e^{-i\omega t}[\xi_{II}e^{in_2\omega(\sigma-L_I)}+ \eta_{II}e^{ -in_2\omega (\sigma-L_I)}],
\end{equation}
where $\xi_I, \eta_I,\xi_{II},\eta_{II}$ are constants. Analogously, using the same relationship $H=\pm nE$ as previously, we have for the magnetic field
\begin{equation}\label{27}
H_I(\sigma,t)= n_1 e^{-i\omega t}\left[ \xi_Ie^{in_1\omega \sigma} -\eta_Ie^{-in_1\omega \sigma }\right],
\end{equation}
\begin{equation}\label{28}
H_{II}(\sigma,t)= n_2 e^{-i\omega t}\left[ \xi_{II}e^{in_2(\sigma-L_I)}-\eta_{II}e^{-in_2\omega (\sigma-L_I)}\right].
\end{equation}
 {We omit the immaterial common time-dependent  factor $e^{-i\omega t}$, which is present in both (\ref{25}),(\ref{26}) for the electric fields and (\ref{27}),(\ref{28}) for magnetic fields (see also \cite{brevik90} )} and introduce the shorthand notation
\begin{equation}
\delta_1=n_1\omega L_I, \quad \delta_2=n_2\omega L_{II}.
\end{equation}
The boundary conditions at the junctions are, for the electric field,
\begin{equation}
-\frac{E_0}{\varepsilon_1}+\xi_Ie^{i\delta_1}+\eta_Ie^{-i\delta_1}=-\frac{E_0}{\varepsilon_2}+\xi_{II}+\eta_{II}, \quad \sigma=L_I, \label{30}
\end{equation}
\begin{equation}
-\frac{E_0}{\varepsilon_2}+\xi_{II}e^{i\delta_2}+\eta_{II}e^{-i\delta_2}= -\frac{E_0}{\varepsilon_1}  +\xi_I+\eta_I, \quad \sigma=L, \label{31}
\end{equation}
and for the magnetic field,
\begin{equation}
n_1(\xi_1e^{i\delta_1}-\eta_Ie^{-i\delta_1})= n_2(\xi_{II}-\eta_{II}), \quad \sigma= L_I,
\end{equation}
\begin{equation}
n_1(\xi_I-\eta_I)= n_2(\xi_{II}e^{i\delta_2}-\eta_{II}e^{-i\delta_2}), \quad \sigma=L,
\end{equation}
where $E_0=\theta_0 B_e= g_{a\gamma\gamma}a_0 B_e$ as before.

{
  Note: similarly as in Ref.~\cite{millar17}, we have here assumed $\omega = \omega_a$. That is, the frequency $\omega$ of the forced oscillations has been taken to be the same as the axion frequency $\omega_a$ associated with the magnetic field ${\bf B}_e$. This is the most natural approach. It is however possible to take a more general approach, in which $\omega$ is assumed to be arbitrary.
It  means that the terms with $E_0$ in (\ref{30}) and (\ref{31}) would need to be corrected with the factor $ e^{i(\omega_a-\omega)t}.$
The reason for taking $\omega$  to be equal to the axion frequency $\omega_a$ is that  in  measurements  the bulk of contribution comes from  values of omega  around $\omega_a$. Indeed the effective/measured   quantities  are always averages over some period of
time, and the mentioned time-dependent oscillations  give   a delta function when
integrated over all  times.  If averaging  over  any  finite period of time,  the contribution  of the regions of $\omega$ substantially different from $\omega_a$ due to the fast oscillating  phase factor add up to zero   (cf. the Lebesgue-Riemann theorem).   We assume that in \cite{millar17} for the same reason also such a value for $\omega$ has been chosen.}\\
We introduce the symbol {  $ n_{12}$} for the refractive index ratio,
{
\begin{equation}
{  n_{12}}=\frac{n_1}{n_2},
\end{equation}}
and consider the scheme
\begin{equation}
\left(\begin{array}{rrrr}
e^{i\delta_1} & e^{-i\delta_1} & -1 & -1 \\
1 & 1 & -e^{i\delta_2} & -e^{-i\delta_2} \\
{  n_{12}} e^{i\delta_1} & -{  n_{12}}e^{-i\delta_1} & -1 & 1 \\
{  n_{12}} & -{  n_{12}} & -e^{i\delta_2} & e^{-i\delta_2}\end{array}
\right)
\left(\begin{array}{c}
\xi_I \\
\eta_I \\
\xi_{II}\\
\eta_{II}
\end{array}
\right)
=
\left(\begin{array}{c}
E_0\left( \frac{1}{\varepsilon_1}-\frac{1}{\varepsilon_2}\right) \\
E_0\left( \frac{1}{\varepsilon_1}-\frac{1}{\varepsilon_2}\right) \\
0 \\
0
\end{array}
\right).
\end{equation}
Here the determinant $D$ of the system matrix $M_{ik}$ can be calculated to be
\begin{equation}
D= \det M_{ik}=-8 {  n_{12}} +2(1+{  n_{12}})^2\cos (\delta_1+\delta_2)-2(1-{  n_{12}})^2\cos (\delta_1-\delta_2). \label{determinant}
\end{equation}}
This is a real quantity. We can now calculate explicit expressions for the field amplitudes $\xi_I, \eta_I, \xi_{II}, \eta_{II}$ ,in the two regions of the string. The expressions become complicated,  and will not be given here. There is one particular case of interest, however,  namely when $n_2$ becomes large in comparison to $n_1$ so that the ratio ${  n_{12}}\rightarrow 0$. The lengths $L_I$ and $L_{II}$ are assumed arbitrary. From Eq.~(\ref{determinant}) we obtain in this case
\begin{equation}
D({ \color{red} n_{12}} \rightarrow 0)= 2\cos(\delta_1+\delta_2)-2\cos(\delta_1-\delta_2)= -4\sin \delta_1\sin\delta_2, \label{37}
\end{equation}
and after some calculation we obtain, as an example, the first of the amplitudes in the region $0<\sigma < L_I$ (note that $1/\varepsilon_2\rightarrow 0$),
\begin{equation}
\xi_I= \frac{E_0}{2\varepsilon_1}\left[ 1-i\frac{1-\cos\delta_1}{\sin\delta_1}\right], \quad {  n_{12}}\rightarrow 0.
\end{equation}
It is noteworthy that this expression does not contain the phase $\delta_2$ related to the length $L_{II}$. It is of further interest to consider the case where the length $L_I\rightarrow 0$, corresponding to a kind of point defect sitting on an otherwise uniform string. As $\delta_1\rightarrow 0$ in this case, we see from the last equation that
\begin{equation}
\xi_I= \frac{E_0}{2\varepsilon_1}, \quad {  n_{12}}\rightarrow 0, L_I \rightarrow 0,
\end{equation}
which is a real quantity. If $E_0=0$,  the axion-induced forced oscillations vanish.

In conclusion, we have managed to show that the axionic electrodynamic scheme is flexible enough to upheld stationary oscillations in the closed string geometry.

\section{Casimir effect for the closed string}
We now put $E_0=0$, so that the forced axion-induced oscillations vanish, and those remaining possible are only the free oscillations. They correspond to the system determinant $D$ being zero. For a given set of geometric quantities $L_I, L_{II}, n_1, n_2$, the eigenfrequencies $\omega$ are thus found by solving  $D=0$ using Eq.~(\ref{determinant}). Moreover, this geometry makes it very natural to consider the Casimir effect also, after applying an appropriate regularization. This will be the topic of the present section. Actually, this situation bears a considerable similarity with the Casimir energy for the relativistic piecewise uniform string, as mentioned earlier \cite{brevik90,li91,elizalde93,brevik96,brevik99,brevik02,bordag12}.\\
{
 For clarity, we ought to clarify the difference between the approach in Sec.~IV and that of the present section in some more detail. The two cases are physically different. In Sec.~IV assuming $E_0 \neq 0$, the system was exposed to forced oscillations, primarily with  frequency $\omega_a$ (eventually with arbitrary $\omega$), giving expressions for the energy $E$ as a function of the refractive index ratio ${ \color{red} n_{12}}=n_1/n_2$.  In the present section the frequencies are no longer up to a choice, but are determined by the determinant condition $D=0$. The Casimir effect is under stationary conditions linked to discrete eigenvalues. As just mentioned, it is this  condition that is just analogous to that  encountered in the study of  the  Casimir effect for the piecewise uniform string.}
Before embarking on the general case, it is convenient to start with special cases of interest. The simplest case is when the string is uniform,  ${ \color{red} n_{12}}=1$, corresponding to $n_1=n_2= n$. Then,
\begin{equation}
D({ \color{red} n_{12}}=1)= -8(1-\cos \omega nL), \label{40}
\end{equation}
so that the eigenfrequencies become
\begin{equation}
\omega_N= \frac{2\pi N}{nL},
\end{equation}
with $N= 1,2,3,...$  Assuming implicitly that we apply a regularization procedure, for instance that of a cutoff regularization, we can express the Casimir energy as
\begin{equation}
E_{\rm uniform}= 2\times \frac{1}{2} \sum_{N=1}^\infty \omega_N,
\end{equation}
where the factor 2 in front accounts for the degeneracy of the left-right running modes.  The result is (cf., for instance, Ref.~\cite{brevik90})
\begin{equation}\label{58}
E_{\rm uniform}=-\frac{\pi}{6L}.
\end{equation}.
The next special case of interest is when ${ \color{red} n_{12}}\rightarrow 0$, the case considered also in the previous section (note that the value of $n_1$ itself is arbitrary, not necessarily small).  Then, from Eq.~(\ref{37}) we see that the eigenfrequencies occur in two branches,
\begin{equation}
\omega_N= \frac{\pi N}{n_1 L_I},
\end{equation}
\begin{equation}
\omega_N= \frac{\pi N}{n_2L_{II}},
\end{equation}
again with $N=1,2,3,...$ (there is no degeneracy in this case, thus no extra factor 2). Introducing the symbol $s$ for the length ratio,
\begin{equation}
s= \frac{L_{II}}{L_I},
\end{equation}
we get in this case
\begin{equation}
E({ \color{red} n_{12}}\rightarrow 0)= -\frac{\pi}{24 L}\left( s+\frac{1}{s}-2\right).
\end{equation}
This Casimir energy is in general negative, but attains its maximum value zero when the pieces have equal lengths, $s=1$.

In the general case when${ \color{red} n_{12}}$ is arbitrary the situation is more complex, but can yet be handled in a reasonably simple  way by making use of { Cauchy's  argument principle}. This principle states that any meromorphic function $g(\omega)$ satisfies the equation
\begin{equation}
\frac{1}{2\pi i}\oint \omega \frac{d}{d\omega}\ln g(\omega)d\omega = \sum \omega_0 -\sum\omega_{\rm poles}, \label{48}
\end{equation}
where $\omega_0$ are the zeros and $\omega_{\rm poles}$ are the poles of $g(\omega)$ inside the contour of integration. This contour is taken to be a semicircle of large radius $R$  in the right half plane. A definite advantage of this method is that the multiplicities of zeros and poles are automatically included. The argument principle was introduced in connection with the Casimir effect by van Kampen {\it et al.} \cite{vankampen68}, and has later been made use of extensively (cf., for instance, Refs.~\cite{brevik02,elizalde93,brevik02,barash89}).

For the function $g(\omega)$ it is natural to start from  the expression for $D$ in Eq.~(\ref{determinant}), but we  have to normalize it in a convenient way. We first  introduce a convenient parametrization which relates the pieces $L_I$ and $L_{II}$ to the total length $L$,
\begin{equation}
L_I= pL, \quad L_{II}=qL, \quad p+q=1.
\end{equation}
Then,  define  the quantity $A$ as
\begin{equation}
A=  \frac{1}{4}(1+{  n_{12}})^2\cos [(n_1 p+n_2 q)\omega L] -\frac{1}{4}(1-{  n_{12}})^2\cos[(n_1p-n_2q)\omega L].
 \end{equation}
For a uniform string, ${  n_{12}}=1~(n_1=n_2=n)$, we have $A=\cos(n\omega L)$. We now  define $g(\omega)$ as
\begin{equation}
g(\omega)= \left| \frac{1-A}{A}\right|.
\end{equation}
With this form, the big semicircle does not contribute to the integration.  Let $\omega=i\xi$, where $\xi$ is the frequency along the imaginary axis. This Wick rotation implies that the quantity $A$ goes into
\begin{equation}
A \rightarrow A(\xi)= \frac{1}{4}(1+{  n_{12}})^2\cosh [(n_1 p+n_2 q)\xi L] -\frac{1}{4}(1-{  n_{12}})^2\cosh[(n_1p-n_2q)\xi L].
\end{equation}
By performing a partial integration along the imaginary axis,  and observing that the positive and negative frequencies contribute equally, we obtain for the Casimir energy
\begin{equation}
E=\frac{1}{2\pi}\int_0^\infty \ln \left| \frac{1-A(\xi)}{A(\xi)}\right|d\xi. \label{68}
\end{equation}
With our adopted  normalization, we can now test the results for $E$ calculated in special cases. To simplify the formalism somewhat, we choose $n_1=1$, the lowest possible value for $n_1$  in nondispersive theory. Then $n_2=1/{  n_{12}}$. Let us also choose $L=1$. This implies that the expression for $A(\xi)$ becomes simpler,
\begin{equation}
A(\xi)= \frac{1}{4}(1+{  n_{12}})^2\cosh \left[\left( p+\frac{ q}{{  n_{12}}}\right)\xi \right] -\frac{1}{4}(1-{  n_{12}})^2\cosh\left[\left(p-\frac{q}{{  n_{12}}}\right)\xi \right].
\end{equation}

\bigskip
\noindent {\it 1. The case of a uniform  string.}

\noindent  In this case $A(\xi)\big|_{x=1} = \cosh \xi$ for all $p$, and we get
\begin{equation}
E_{\rm uniform}= \frac{1}{2\pi}\int_0^\infty \ln \left| \frac{1-\cosh \xi}{\cosh\xi}\right| d\xi.
\end{equation}
The integral can be calculated analytically by means of polylogarithm functions and is equal to
\begin{equation}\label{70}
E_{\rm uniform}=-3 \frac{\pi}{16}.	
\end{equation}
It is worth noticing it is different from (\ref{58}) by a multiplicative factor $\frac{9}{8}$. One could think that this is  an inconsistency as the calculations seem to refer to the same Casimir energy $ E$  for the homogenous string. Actually, the two situations are different:\\The energy of Eq.~(\ref{58}) is calculated for a scalar field, so quantum numbers run  from $N=1$ to infinity , while Eq.~(\ref{70}) is calculated for the electromagnetic field, where $N=0$ contributions are not zero.
It is known in electromagnetic Casimir problems  for dielectrics
that both TE and TM polarizations contribute with zero frequency \cite{milton01}. Although the two cases are not directly comparable, we note
 that the result of the integral (\ref{70}) is obtained from
\begin{equation}
E_{\rm uniform}=-\frac{1}{2\pi}(4 \Li_2(1)-\Li_2(-1)),	
\end{equation}
where the first term in brackets gives the  energy (\ref{58}), and the second term  gives the difference.

%
%
%

\noindent {\it 2. The case when $p=1/2.$}

\noindent The string is then divided into two equal halves, the refractive index ratio  ${ \color{red} n_{12}}$ being arbitrary. Now
\begin{equation} \label{58li}
A(\xi)\big|_{p=1/2}= \frac{1}{4}(1+{  n_{12}})^2\cosh \left[ \left( 1+\frac{1}{{  n_{12}}}\right) \frac{\xi}{2}\right] -\frac{1}{4}(1-{  n_{12}})^2\cosh \left[ \left( 1-\frac{1}{{  n_{12}}}\right)\frac{\xi}{2}\right],
\end{equation}
and the Casimir energy is found by inserting this expression into Eq.~(\ref{68}).
\begin{figure}[h]\label{fig:2}
	\includegraphics[scale=0.6]{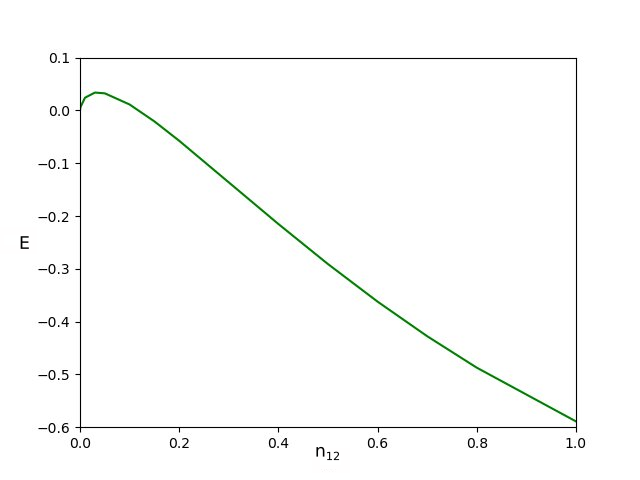}
	\caption{Graph of E versus ${  n_{12}}$ in the case 2, where   ${  n_{12}}$ is restricted to the region $0< {  n_{12}}<1$. We show that E can be  positive or negative depending on the value of   ${  n_{12}}$: for $ {  n_{12}} \lessapprox 0.15$, E is positive and has a maximum at $ {  n_{12}} \approx 0.04$, for bigger  ${  n_{12}}$ the energy is negative, until it reaches the value in Eq.~(\ref{70}) for  ${  n_{12}}=1$.}
\end{figure}
It is not trivial to find an analytical solution for the integral (\ref{68}),   so we calculated it numerically by a Python program. We show the resulting plot in Fig. 2.
\\ It is worth noticing how the energy $E$ is positive or negative depending on the value of ${  n_{12}}$.  For ${  n_{12}} \lessapprox 0.15$, $E$ is positive and has a maximum at ${  n_{12}} \approx 0.04$; for larger ${  n_{12}}$ the energy is negative until it reaches the value (\ref{70}) for ${  n_{12}}=1$. In the following case for $ p\rightarrow 0$ we are discussing the similarity with this case.
\begin{figure}[h]\label{fig:3}
	\includegraphics[scale=0.6]{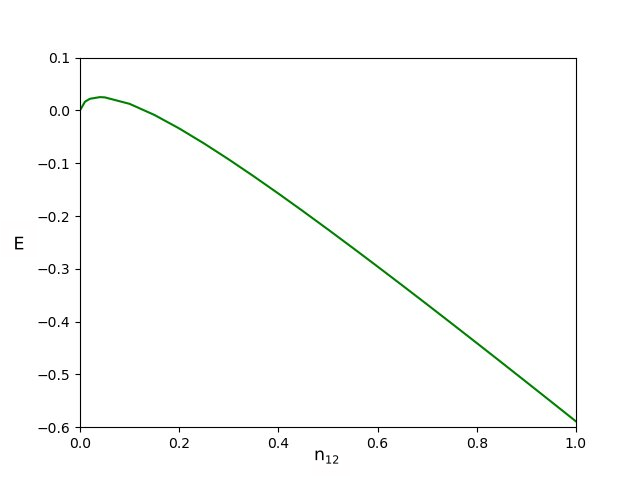}
	\caption{Graph of $E$ versus ${  n_{12}}$ in the case 3, analogous to Fig. 2.  We show that  $E$ is positive or negative depending on  the value of ${  n_{12}}$: for ${  n_{12}} \lessapprox 0.15$, $E$ is positive and has a maximum at ${  n_{12}} \lessapprox 0.04$; for larger ${  n_{12}}$, the  energy is negative until it reaches  Eq.~(\ref{70}) for ${  n_{12}}=1$. The main difference from Fig. 2  relies on the behaviour of the function in the middle region, where the energy of case 3 diminishes  more slowly than in the case 2.}
\end{figure}

\noindent {\it 3. The case when $ p\rightarrow 0$.}

\noindent This case is of interest since it corresponds to a "particle" sitting on a uniform string. We have now $A(\xi)\big|_{p\rightarrow 0}= {  n_{12}}\cosh (\xi/{  n_{12}})$, and the Casimir energy becomes
\begin{equation}\label{59}
E= \frac{1}{2\pi}\int_0^\infty \ln \left|\frac{1-{  n_{12}}\cosh (\xi/{  n_{12}})}{{  n_{12}}\cosh (\xi/{  n_{12}})}\right|d\xi.
\end{equation}
{
As before, we  assume that ${  n_{12}}$ lies in the interval $0<{  n_{12}}<1$. Figure 3 shows the same behavior except from the middle region   $0.15 \lessapprox {  n_{12}}< 1$, where the energy of case 3 diminishes  slower than in case 2.
These similarities in the extremes for ${  n_{12}} \lessapprox 0.15$ and ${  n_{12}}=1$ can be understood. It is easy to verify that the integrals in Eqs.~(\ref{58li}) and (\ref{59}) give the same result for ${  n_{12}}=0$, because we have no fields due to boundary conditions and the medium in region II behaving as a perfect mirror so we get $E=0$. The same is valid for ${  n_{12}}=1$, as the cases 1, 2 and 3 are equivalent when the string is uniform.}

\section{Curvature-induced enhancement of the produced electric field at a dielectric boundary}

As our last topic, we will return to the cylindrical haloscope idea, emphasizing a point that to our knowledge has not been considered before. Namely, the extra produced electric field $E^\gamma$ occurring in the interior cylindric vacuum region will necessarily be enhanced near the centre of the cylinder. The enhancement is solely caused by the curvilinear geometry. Obviously, this is a point that can be  of experimental interest. We assume the case when   the optical frequency is equal to the axion frequency $\omega_a$.


As before, assume there is an infinitely  long cylindrical vacuum region with radius $R$, surrounded by an exterior massive dielectric  environment of high refractive index. There is moreover   a strong uniform static magnetic field ${\bf B}_e$  in the $z$ direction. We assume there are only the time-varying axions present in the system, together with the extra photons $E^\gamma$ that they generate at the boundary.  The solutions of the Bessel equations in the exterior and interior regions are cylinder functions. No propagation of electromagnetic modes are assumed to take place in the $z$ direction, and we can start from   the usual electromagnetic mode  expressions for the axial electric and the azimuthal magnetic field components in the exterior and interior regions (cf., Ref.~\cite{stratton41}, page 525)
\begin{equation}
E_z^{\rm ext}=H_0^{(1)}(k_2r)a_{\rm ext}e^{-i\omega t}, \quad H_\theta^{\rm ext}= in_2 H_0^{(1)'}(k_2r)a_{\rm ext}e^{-i\omega t}, \label{52}
\end{equation}
\begin{equation}
E_z^{\rm int}=J_0(\omega r)a_{\rm int}e^{-i\omega t}, \quad H_{\theta}^{\rm int}= iJ_0'(\omega r)a_{\rm int}e^{-i\omega t}. \label{53}
\end{equation}
{
Here $k_1$ is the wave number in the inner region $0<r<R$, and $k_2$ refers to the outer region $r>R$.
 $H_p^{(1)}$  is the  Hankel functions of the first  kind, of order $p$.  Azimuthal symmetry is assumed, so that only $p=0$ applies. Outgoing waves are assumed on the outside,  and on the inside a stationary wave field is assumed, finite at the center.

In the present case the situation is different,  as the antenna-like property of the boundary $r=R$ causes radiation to occur  into the inward also. It is illustrative here to give some of the mathematical background \cite{stratton41}. In general, the complex Hankel functions of the first and the second kind are defined as
\begin{equation}
H_p^{(1)}(\rho)= J_p(\rho)+iN_p(\rho), \quad  H_p^{(2)}(\rho)= J_p(\rho)-iN_p(\rho),
\end{equation}
where $J_p(\rho)$ and $N_p(\rho)$ are the real Bessel and Neumann functions. For large arguments $\rho \gg 1$ one has asymptotically, after multiplying with the common time factor $e^{-i\omega t}$,
\begin{equation}
H_p^{(1)}(\rho)= \sqrt{\frac{2}{\pi\rho}}\,e^{i(\rho -\frac{2p+1}{4}\pi  -i\omega t)},
 \quad  H_p^{(2)}(\rho)= \sqrt{\frac{2}{\pi\rho}}\,e^{-i(\rho -\frac{2p+1}{4}\pi  + i\omega t)}. \label{approx}
\end{equation}
This shows that $H_0^{(2)}$ is the correct cylinder function to use in the present case. It describes waves produced at the boundary $r=R$, moving inward.

 Now replacing  $J_0$  by  $H_0^{(2)}$ in  Eq.~(\ref{53}), the equation takes the form
\begin{equation}
E_z^{\rm int}=H_0^{(2)}(\omega r)a_{\rm int}e^{-i\omega t}, \quad H_{\theta}^{\rm int}= iH_0^{(2)'}(\omega r)a_{\rm int}e^{-i\omega t}.
\end{equation}}
As before, $k_2= n_2\omega$ with $n_2$ real.
The coefficients $a_{\rm ext}$ and $a_{\rm int}$ are determined by the boundary conditions at $r=R$. Looking back at the treatment in Sec.~III, we will require the correspondence
\begin{equation}
E_2^\gamma \rightarrow H_0^{(1)} (k_2R)a_{\rm ext}e^{-i\omega t}, \quad E_1^{\gamma} \rightarrow H_0^{(2)}(\omega R)a_{\rm int}e^{-i\omega t}.
\end{equation}
We assume in the following that the radius $R$ is so large that the approximate versions (\ref{approx}) for the Hankel functions are applicable on the boundary.
As $H_0^{(1)'}(\rho) = iH_0^{(1)}(\rho)$, \, $H_0^{(2)'}(\rho)= -iH_0^{(2)}(\rho)$, we can write the boundary conditions as
\begin{equation}
H_0^{(1)}(k_2R)a_{\rm ext}-\frac{E_0}{\varepsilon_2}= H_0^{(2)}(\omega R)a_{\rm int}-E_0,
\end{equation}
\begin{equation}
n_2 H_0^{(1)}(k_2R)a_{\rm ext}=-H_0^{(2)}(\omega R) a_{\rm int},
\end{equation}
where $E_0 = \theta_0B_e$, as before.
 The  coefficients  $a_{ \rm ext}$ and $a_{\rm int}$ can then be found, and we can express the result as
\begin{equation}
H_0^{(1)} (k_2R)a_{\rm ext} = -\frac{E_0}{n_2}\left( 1-\frac{1}{n_2}\right),
\end{equation}
\begin{equation}
H_0^{(2)}(\omega R)a_{\rm int}= E_0\left( 1-\frac{1}{n_2}\right),
\end{equation}
in agreement with Eq.~(\ref{check}) obtained for plane geometry.

We now move on to the central point in this analysis, which is to observe the behavior of the inward-emitted electric field $E_z$ near the center of the cylinder. As the magnitude of $H_0^{(2)}(\rho)$  increases logarithmically as $\rho \rightarrow 0$, there will necessarily be an enhancement of the field in this region. {  (Note that this enhancement is a pure focusing  effect caused by the boundary $r=R$, and is not related to a divergence of the source.)}
   One might    argue that the accumulation of electric fields near $r=0$ would lead to an unstable situation. We may avoid this, at least conceptually, by assuming there to be a perfect cylindrical absorber of small radius $r=\delta$ centered at the $z$ axis, thus allowing for stable conditions.

Letting the  inward-generated field at $r=R$ be represented by $H_0^{(2)}(\omega R)$, we may for small values of the argument $\rho$ make use of the approximation
\begin{equation}
H_0^{(2)}(\rho)= J_0(\rho)-iN_0(\rho)= 1+\frac{2i}{\pi}\ln \frac{2}{\gamma \rho}, \quad \rho \ll 1, \label{lowenergy}
\end{equation}
with $\gamma = 1.78107$.
We need only consider the magnitudes of the field expressions. Thus the magnitudes of the fields measured at the minimum radius $ \delta$ and at the cylinder radius $R$ are related by
\begin{equation}
\left|\frac{E_z^\gamma(\delta)}{E_z^\gamma(R)}\right|=           \left|\frac{H_0^{(2)}(\omega \delta/c)}{H_0^{(2)}(\omega R/c)}\right|=
\left| \frac{1+\frac{2i}{\pi}\ln \frac{2c}{\gamma \omega \delta}}{H_0^{(2)}( \omega R/c)}\right| \label{ratio}
\end{equation}
in physical units.

We make finally some numerical estimates. Take $R=20~$cm as a reasonable value for the cylinder radius, and choose for definiteness $m_a c^2= 10^{-4}~$eV for the axion energy, what corresponds to $\omega= 1.52\times 10^{11}~$rad/s. Then $\omega R/c=1.0\times 10^2$, thus justifying use of the approximation (\ref{approx}). We get $|H_0^{(2)}(\omega R/c)|= 0.080~$. With the minimum radius chosen as $\delta = 100~\mu$m, corresponding to $\rho_{\rm min}= \omega \delta/c= 0.051$, it is seen that also the low-argument approximation (\ref{lowenergy}) is justified. We obtain for the ratio (\ref{ratio})
\begin{equation}
\left|\frac{E_z^\gamma(\delta)}{E_z^\gamma(R)}\right|= 12.5\times |1+2.0\, i|  =28.0.
\end{equation}
There occurs thus a considerable enhancement of the amplitude in the neighbourhood of the cylinder center.  While the numerical values above were chosen so as to satisfy the mathematical approximations, there is clearly a flexibility of choosing different parameters that can better fit experimental conditions.

We emphasize again that this enhancement is solely a geometrical focusing effect, being an extension of the theory worked out earlier \cite{millar17} for plane geometry. An observation of the increased signal near $r=0$ may be of experimental interest. The idea may be looked upon as an alternative to the idea recently put forward in Ref.~\cite{liu22}, concerning  the broadband solenoidal haloscope.

The situation discussed above is obviously  an  idealistic  one. In practice, an accurate positioning
of the absorber would be  needed, and the impedance matched absorber should be with  the property
of  having  no reflection. One may ask if there exist broadband frequency reflection-less absorbers in the THz region that might be suitable for this purpose.  Perhaps the use of metamaterials turn out to be useful in this context.  We may mention here  the  paper of Zhang {\it et al.} on a broadband THz absorber based on dispersion-engineered catenary coupling in dual metasurface \cite{zhang19}. Indeed the authors of \cite{zhang19} were able to obtain an ultra-broadband THz absorber with a bandwidth range from 0.52 to 4.4 THz, thus just in the region of interest. Similar investigations, in the GHz region, are given by Chen {\it et al.} \cite{chen18}, and Chen {\it et al.} \cite{chen17}. A general review on metamaterial absorbers is recently given by Abdulkarim {\it et al.} \cite{abdulkarim22}.

\section*{Acknowledgments}
We thank the Erasmus project, the Norwegian University of Science and Technology and the University of Palermo for the support of the Erasmus Traineeship for  M.Sc. Amedeo Maria Favitta. Part of this work was performed in the context of his Master Thesis.

\end{document}